# Superconductor-to-Metal Quantum Phase Transition in Overdoped $La_{2-x}Sr_xCuO_4$

T.R. Lemberger,[1,*] I. Hetel,[1] A. Tsukada,[2] M. Naito[2], and M. Randeria[1]

[1]*Department of Physics, The Ohio State University, Columbus, Ohio, 43210, USA*

[2]*Tokyo University of Agriculture and Technology, Tokyo, Japan*

We investigate $T_c$ and magnetic penetration depth $\lambda(T)$ near the superconductor-metal quantum phase transition in overdoped $La_{2-x}Sr_xCuO_4$ films. Both $T_c$ and superfluid density $n_s$, $\propto \lambda^{-2}$, decrease with overdoping. They obey the scaling relation $T_c \propto [\lambda^{-2}(0)]^{\alpha}$ with $\alpha \approx ½$. We discuss this result in the frameworks of disordered *d*-wave superconductors and of scaling near quantum critical points. Our result, and the linear scaling ($\alpha \approx 1$) found for the more anisotropic $T\ell_2Ba_2CuO_{6+\delta}$, can both be understood in terms of quantum critical scaling, with different dimensionalities for fluctuations.

*Corresponding Author: Thomas R. Lemberger, Lemberger.1@osu.edu, (614) 292-7799, (614) 292-7557 (FAX).

The superconductor-to-nonsuperconductor transitions in cuprates, as functions of carrier concentration, give insights into quantum phase transitions (QPT's) in general [1] and into the phenomenon of high temperature superconductivity in particular [2-4]. On the underdoped side, the transition is from superconducting to an insulating state with a (pseudo)gap in the electronic excitation spectrum that remains finite through the QPT. The fundamental physics on the overdoped side of the phase diagram is profoundly different because there is no pseudogap, the superconducting gap becomes progressively smaller [5] with doping, and the QPT is from superconductor to a metal that looks like a conventional Fermi liquid in many respects.

Key issues are: (i) Are these transitions first-order or are they quantum critical points (QCP) where quantum fluctuations of the order parameter are important? (ii) How does the presence or absence of an energy gap impact the transitions? (iii) Are these transitions driven by a collapse of the pairing amplitude or by fluctuations of the phase of the superconducting order parameter? (iv) Does anisotropy (*c*-axis *vs. ab*-plane) affect the dimensionality of the QCP's?

$T_c$ as a function of hole doping $p$ takes a quasi-universal form [6], with superconductivity existing for $0.3 \lesssim p \lesssim 0.30$, with a maximum at $p \approx 0.15$, independent of the maximum value of $T_c$ or of *c vs*. *ab*-plane anisotropy. Thus, one might expect a common explanation for the over- and underdoped quantum phase transitions in different compounds.

An early study of several underdoped cuprate compounds suggested that $T_c$ and superfluid density [$n_s \propto \lambda^{-2}$, $\lambda$ = magnetic penetration depth] might be linearly proportional: $T_c \propto \lambda^{-2}(0)$, with a universal slope as critical underdoping is approached [7]. This linear scaling led to the widely accepted view that classical thermal phase fluctuations destroy superconductivity [4] in underdoped cuprates when the superfluid density becomes small, even as the energy gap remains intact. This long-standing view was overturned recently by measurements on severely underdoped $YBa_2Cu_3O_{7-\delta}$ (YBCO) films [8] and crystals [9] showing that scaling is actually sublinear: $T_c \propto [\lambda^{-2}(0)]^{\alpha}$ with $\alpha \approx 0.5$. Sublinear scaling, together with the absence of critical thermal fluctuations near $T_c$, pointed to a 3D QCP [2,3]. The QCP hypothesis was put to a stringent test in a study of two-unit-cell-thick underdoped YBCO films that were 2D by construction. Indeed, linear scaling expected near a 2D QCP was observed for these ultrathin films [10].

The present work focuses on the *overdoped* QPT in $La_{2-x}Sr_xCuO_4$ (LSCO). While the underdoped regime has been explored in several materials, studies in the overdoped regime have focused largely on a single material: $T\ell_2Ba_2CuO_{6+\delta}$ (T$\ell$2201) [11,12]. We are motivated to study LSCO because it can be doped through both over- and underdoped quantum phase transitions, and it is much less anisotropic than T$\ell$2201, thus allowing us to address the key question of the effective dimensionality of fluctuations. Our main results are:

(1) All overdoped samples with high Sr concentrations, $x > 0.22$, have sharp superconducting-to-normal thermal phase transitions, as narrow as 200 mK near the QPT. This suggests that the overdoped QPT is not dominated by inhomogeneity or phase separation.

(2) Near the overdoped QPT, we find sublinear scaling $T_c \propto [\lambda^{-2}(0)]^{\alpha}$ with $\alpha \approx 0.5$ for LSCO, in contrast to the linear scaling ($\alpha \approx 1.0$) seen in T$\ell$2201 [11,12].

(3) We argue that scaling with $\alpha \approx 0.5$ is consistent with either (a) a mean-field QPT driven by gap collapse in a disordered *d*-wave superconductor, or (b) a 3D quantum critical point (QCP). In case (a), asymptotically close to the QPT one must take into account critical fluctuations. Case (b) permits us to reconcile the square-root scaling in LSCO with the linear scaling in T$\ell$2201.

Our $La_{2-x}Sr_xCuO_4$ films were grown by MBE on (001) $LaSrAlO_4$ (LSAO) substrates [13] (see Table I). The films' *c*-axes are perpendicular to the substrate. Compressive strain due to a 0.6% lattice mismatch (*a* = 3.754 Å for LSAO and 3.777 Å for LSCO) gives our films a maximum $T_c$ (≈ 44 K) several K above the maximum $T_c$ of LSCO crystals. Sr doping values are nominal. They are set by atomic beam fluxes during deposition. Well after the first series of films ($d$ = 45 nm) was grown, we decided to grow two additional films, at $x$ = 0.27 and 0.30 ($d$ = 90 nm; last two rows of Table I). These films have somewhat higher resistivities, $T_c$'s and superfluid densities than for the first series, possibly due to a different oxygen vacancy concentration.

Two samples were grown simultaneously at each Sr concentration, one on a narrow substrate for measuring resistivity and the other on a 10×10×0.35 $mm^3$ substrate for measuring $\lambda^{-2}$ . Sheet conductivity, $\sigma d = \sigma_1 d - i\sigma_2 d$, was measured with a low-frequency ($\omega/2\pi$ = 50 kHz) two-coil mutual inductance technique, with drive and pickup coils on opposite sides of the film [14]. Near $T_c$, the real part of the conductivity $\sigma_1(T)$ has a peak that probes the spatial homogeneity of $T_c$. The imaginary part, $\sigma_2(T)$, yields the magnetic penetration depth $\lambda$ *via*: $\lambda^{-2}(T) \equiv \mu_0\omega\sigma_2(T)$. $\lambda^{-2}(T)$ is often loosely referred to as "superfluid density", $n_s$, since the two are proportional.

The *ab*-plane resistivities of our films decrease smoothly with doping, Fig. 1, achieving a low residual resistivity of about 40 μΩ cm at the highest doping, comparable to that of a similarly overdoped LSCO crystal [15]. "$T_c$" defined from where $\rho_{ab}$ vanishes, agrees within a degree or so with "$T_c$" defined from where superfluid appears.

Figures 2 and 3 show $\lambda^{-2}(T)$ for representative LSCO films, illustrating the qualitative feature that $\lambda^{-2}(T)$ for overdoped films has less downward curvature than for under- and optimally-doped films. The same qualitative effect is seen in LSCO powders [16] and in T$\ell$2201 powders [11,12,17]. $\sigma_1$ is plotted as $\mu_0\omega\sigma_1$, ($\mu_0$ = permeability of vacuum = $4\pi\times10^{-7}$ H/m), to facilitate quantitative comparison with $\lambda^{-2} \equiv \mu_0\omega\sigma_2$. Peaks in $\sigma_1(T)$ (Figs. 2 and 3) probe film homogeneity. Films with $x \leq 0.21$ have peaks with structure indicating the presence of several closely-spaced $T_c$'s over the *mm*-scale area probed. On the other hand, films with $x \geq 0.24$ have single peaks about 1 K wide, *e.g.*, the $x = 0.30$ film in Fig. 2. The peak is only 0.2 K wide for the film closest to the QPT (Fig. 3), consistent with good film homogeneity, although there are other experiments [17] suggesting a phase-separated overdoped superconducting state.

It is important to establish the quality of our films by comparison with the literature. $T_c$ vs. $x$ for films (black squares in Fig. 4) follows the usual path, peaking at $x \approx 0.15$ and heading toward zero at $x \approx 0.03$ and 0.30. $\lambda^{-2}(0)$ vs. $x$ for films (red squares in Fig. 4) tracks that of LSCO powders (green circles) [18] up to $x \approx 0.18$. $\lambda^{-2}(0)$ of our films decreases with strong overdoping, with a peak near $x \approx 0.19$, consistent with overdoped T$\ell$2201[11,12] and other LSCO films [19]. Resistivity and superfluid density measurements show that our films are essentially of the same quality as bulk samples.

A detailed examination of how the magnitude and T-dependence of superfluid density change across the phase diagram is presented elsewhere [20]. Here we focus on scaling between $T_c$ vs. $\lambda^{-2}(0)$, Fig. 5. Data from our under- and overdoped LSCO films are shown as open and filled red squares, respectively. Data on other cuprates [8,10] are shown for comparison (see caption). The solid gray line representing square-root scaling is drawn through the underdoped Ca-YBCO thick film data, but it is close to the underdoped LSCO data, too. The solid red line representing square root scaling is drawn through the data for strongly overdoped LSCO films. The light blue line representing linear scaling is drawn through the data for two-unit-cell thick Ca-doped YBCO.

Let us ask *why* $n_s$ decreases with overdoping, even though the carrier density increases. The most natural explanation is the interplay between disorder (scattering rate $1/\tau$) and a *d*-wave pairing interaction, and thus gap $\Delta_0$, that weakens with overdoping. In a disordered *d*-wave superconductor, a simple sum-rule argument suggests a linear suppression of $n_s(0)$ to zero with increasing $1/\Delta_0\tau$, borne out by detailed calculations [21]. In addition, the dirty *d*-wave $T_c$ exhibits a square-root suppression to zero with $1/\Delta_0\tau$ [21], so that: $T_c \sim [\lambda^{-2}(0)]^{1/2}$. We note this is a mean-field result, and for doping close enough to the QPT, the superfluid density necessarily becomes so small that quantum phase fluctuations dominate the physics. It is not known where the crossover to this asymptotic behavior occurs.

Well-known scaling arguments predict [1-3] that $T_c \sim [\lambda^{-2}(0)]^{\alpha}$ near a QCP, with exponent $\alpha \equiv z_Q/(z_Q + D - 2)$, where D = dimensionality and $z_Q$ = quantum dynamical exponent. Since $z_Q$ should not be less than unity, in D = 3 the smallest reasonable exponent is $\alpha = ½$, which is coincidentally the same as the dirty-*d*-wave mean field result. This describes the observed nonlinear scaling in overdoped LSCO reasonably well.

The QCP interpretation also permits us to understand the linear scaling in overdoped T$\ell$2201 [11,12]. It is reasonable to expect 2D fluctuations in T$\ell$2201, which is much more anisotropic than LSCO. For D = 2, the exponent $\alpha = 1$, independent of $z_Q$. Thus the different scalings seen in LSCO and T$\ell$2201 can be attributed to the different dimensionalities of the fluctuations.

It is worth noting that there is experimental evidence for significant interlayer coupling in overdoped LSCO. Ironically, this evidence comes from a 2D, Kosterlitz-Thouless-Berezinski-like transition seen in the most overdoped film, *i.e.*, the abrupt downturn in $\lambda^{-2}$ near the intersection of the KTB line with $\lambda^{-2}(T)$ (Fig. 3). The slope of the KTB line in Fig. 3 is calculated assuming that the film fluctuates as a single 2D entity. For independently fluctuating layers, the slope of the KTB line would be 70 times larger. Analogous features appear in microwave measurements of σ in *underdoped* LSCO films [22], indicating significant interlayer coupling across the LSCO phase diagram. Finally, similar evidence for interlayer coupling is found in "thick" underdoped YBCO films, which also show 3D critical scaling [23].

In summary, we observe in overdoped LSCO that superconductivity diminishes with $T_c \sim [\lambda^{-2}(0)]^{1/2}$. Taken by itself, this behavior may be viewed as a consequence of a mean-field gap collapse in a disordered *d*-wave superconductor. On the other hand, if one seeks a common explanation for the nonlinear scaling in LSCO and the linear scaling in Tℓ2201, then one is led to the interpretation that the scaling observed for strongly overdoped samples is due to 3D and 2D quantum critical points, respectively. The difference in dimensionality would be due to the much higher anisotropy of Tℓ2201. Finally, in moderately *underdoped* LSCO we observe nonlinear $T_c$ vs. $\lambda^{-2}(0)$ scaling that is quantitatively similar to that of underdoped YBCO. However, data on severely underdoped LSCO samples are needed to establish a 3D QCP on the underdoped side.

ACKNOWLEDGEMENTS: This work was supported in part by NSF DMR grant 0203739 (IH), DOE grant FG02-08ER46533 (TRL) and NSF-DMR 0706203 (MR). We acknowledge useful conversations with Ilya Vekhter and David Stroud.

REFERENCES

*Lemberger.1@osu.edu.

[1] S. Sachdev, *Quantum Phase Transitions*, (Cambridge University Press, Cambridge, 1999).

[2] A. Kopp and S. Chakravarty, Nature Phys. 1, 53 (2005).

[3] M. Franz and P. Iyengar, Phys. Rev. Lett. 96, 047007 (2006).

[4] V. J. Emery and S. A. Kivelson, Nature 374, 434 (1995).

[5] A. Ino et al., Phys. Rev. B 65, 094504 (2002).

[6] M.R. Presland et al., Physica (Amsterdam) 176C, 95 (1991).

[7] Y.J. Uemura et al., Phys. Rev. Lett. 62, 2317 (1989).

[8] Y. Zuev, M.-S. Kim, and T.R. Lemberger, Phys. Rev. Lett. 95, 137002 (2005).

[9] D. M. Broun et al., Phys. Rev. Lett. 99, 237003 (2007).

[10] I. Hetel, T.R. Lemberger and M. Randeria, Nature Phys. 3, 700 (2008).

[11] Y.J. Uemura et al., Nature 364, 605-607 (1993).

[12] Ch. Niedermayer et al., Phys. Rev. Lett. 71, 1764 (1993).

[13] M. Naito and H. Sato, Appl. Phys. Lett. 67, 2557 (1995); H. Sato and M. Naito, Physica C 274, 221 (1997); H. Sato et al., Phys. Rev. B 61, 12447 (2000); H. Sato, Physica C 468, 2366 (2008).

[14] S.J. Turneaure, E.R. Ulm, and T.R. Lemberger, J. Appl. Phys. 79, 4221 (1996); S.J. Turneaure, A.A. Pesetski, and T.R. Lemberger, J. Appl. Phys. 83, 4334 (1998).

[15] S. Nakamae et al., Phys. Rev. B 68, 100502(R) (2003).

[16] C. Panagopoulos et al., Phys. Rev. B 60, 14617 (1999).

[17] Y.J. Uemura, Solid State Comm. 120, 347 (2001).

[18] C. Panagopoulos et al., Phys. Rev. B 67, 220502 (2003).

[19] J.-P. Locquet et al., Phys. Rev. B 54, 7481 (1996).

[20] T.R. Lemberger et al., arXiv:1008.2744.

[21] E. Puchkaryov and K. Maki, Eur. J. Phys. B **4**, 191 (1998); Y. Sun and K. Maki, Phys. Rev B **51**, 6059 (1995).

[22] H. Kitano et al., Phys. Rev. B 73, 092504 (2006).

[23] Y.L. Zuev et al., Physica C 468, 276 (2008).

TABLE I. Properties of $La_{2-x}Sr_xCuO_4$ films grown by MBE on LSAO (100). "*x*" values are nominal. The last two films were grown well after the others with a slightly different protocol, and they are twice as thick.

| *x* | $T_c(\rho_{ab})$ (K) | $T_c(\lambda^{-2})$ (K) | $\lambda^{-2}(0)$ ($\mu m^{-2}$) | $\rho_{ab}$ (50K) ($\mu\Omega$ cm) |
|---|---|---|---|---|
| 0.06 | 17.5 | 16 | 1.3 | 590 |
| 0.06 | 24 | 23 | 4.5 | 377 |
| 0.09 | 39.7 | 33 | 6 | 170 |
| 0.09 | 40 | 38 | 10.5 | 160 |
| 0.12 | 40.1 | 39 | 12.5 | 140 |
| 0.15 | 44 | 42 | 17.4 | 90 |
| 0.18 | 41 | 38 | 21.5 | 54 |
| 0.21 | 33 | 32 | 20.3 | 48 |
| 0.24 | 19 | 18.5 | 11.1 | 37 |
| 0.27 | 4.0 | 3.9 | 0.15 | 31 |
| 0.27 | 21 | 20 | 3.4 | 70 |
| 0.30 | 9 | 8.5 | 0.8 | 56 |

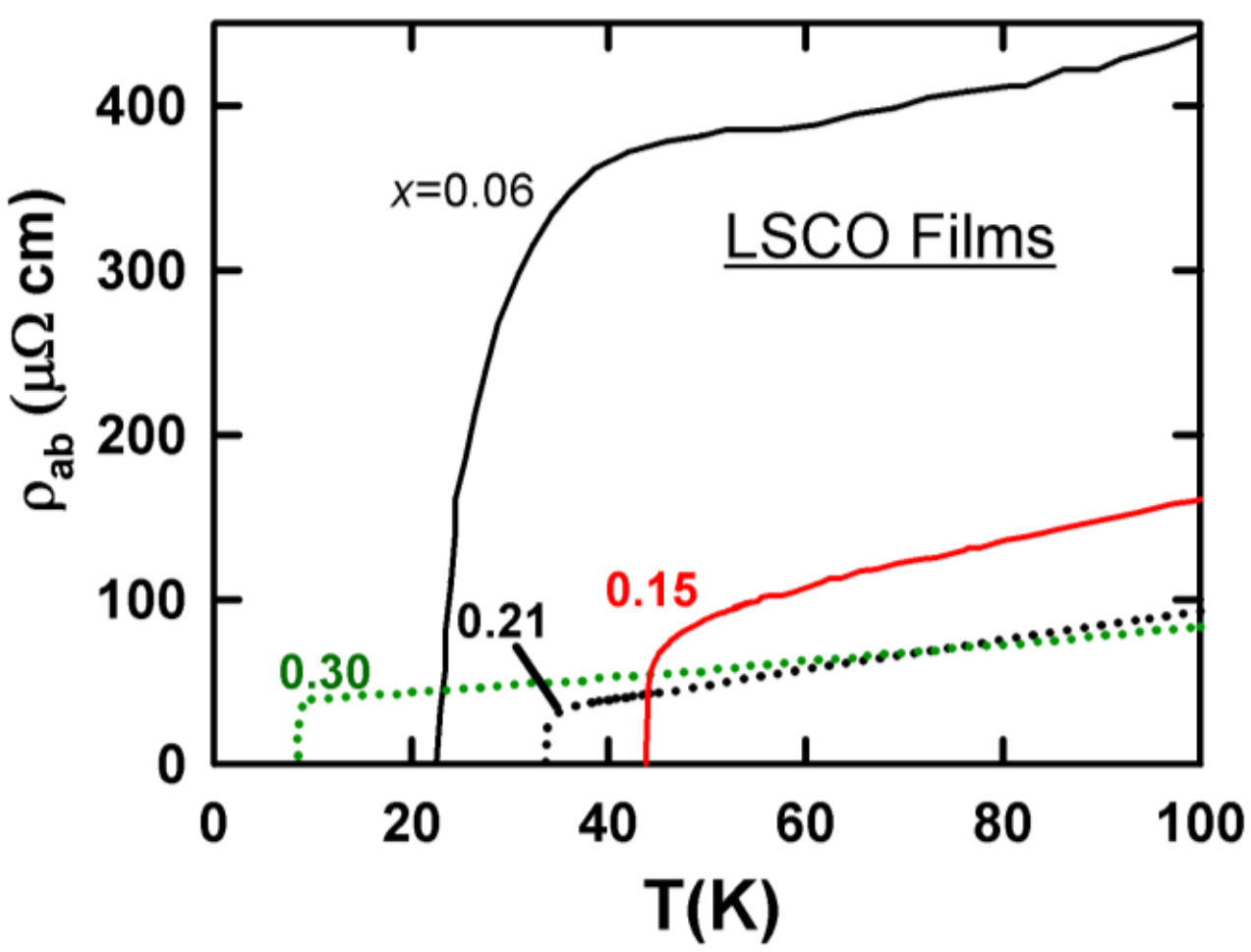

FIG. 1 (color online). *ab*-plane resistivity $\rho_{ab}(T)$ for typical $La_{2-x}Sr_xCuO_4$ films: $x = 0.06, 0.15, 0.21, 0.30$, illustrating the shallow minimum in resistivity and the maximum in $T_c$ as functions of doping.

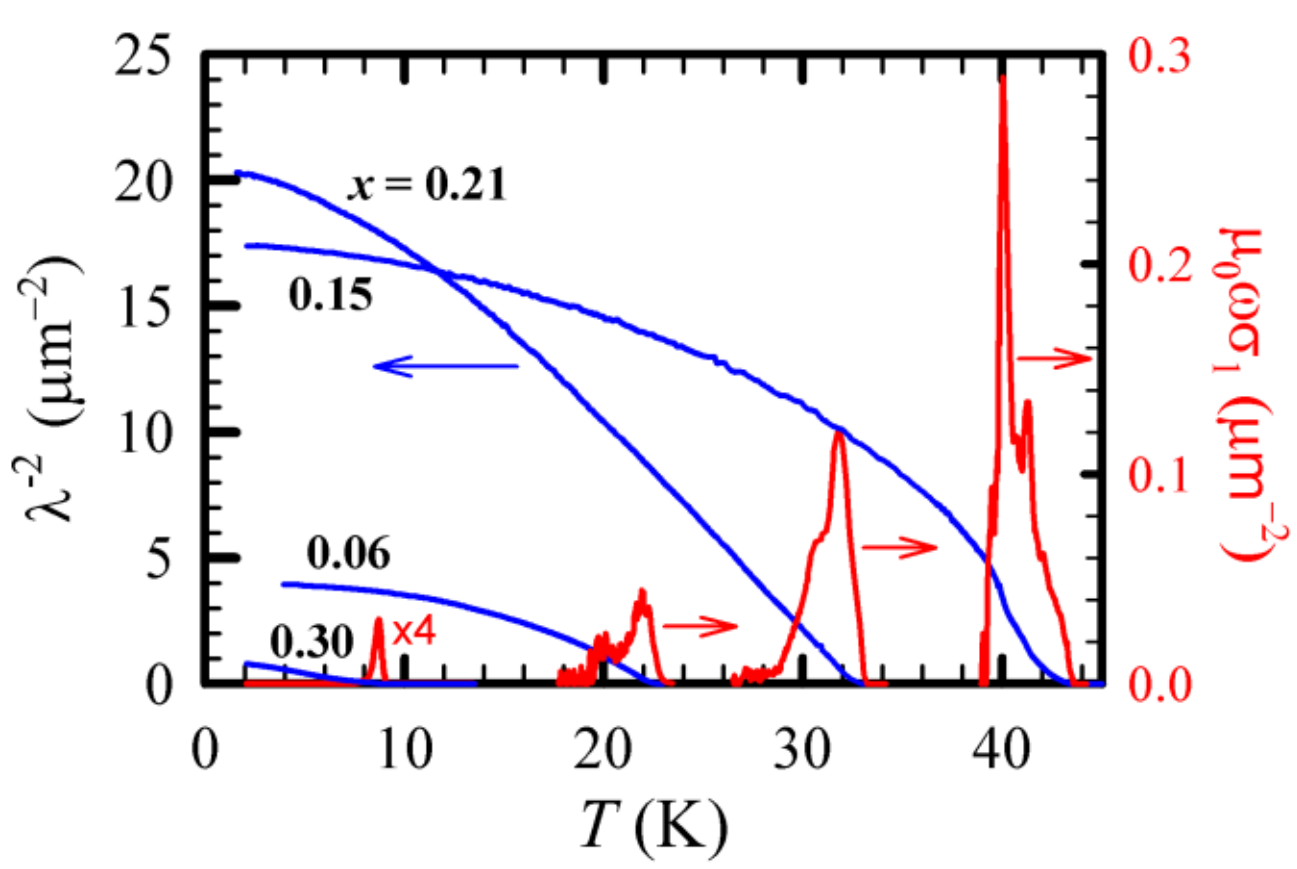

FIG. 2 (color online). $\lambda^{-2} \equiv \mu_0\omega\sigma_2$ (blue curves) and $\mu_0\omega\sigma_1(T)$ (red peaks) measured at $\omega/2\pi = 50$ kHz for $La_{2-x}Sr_xCuO_4$ films: $x = 0.06, 0.15, 0.21, 0.30$, illustrating the maxima in $T_c$ and $\lambda^{-2}(0)$ as functions of doping.

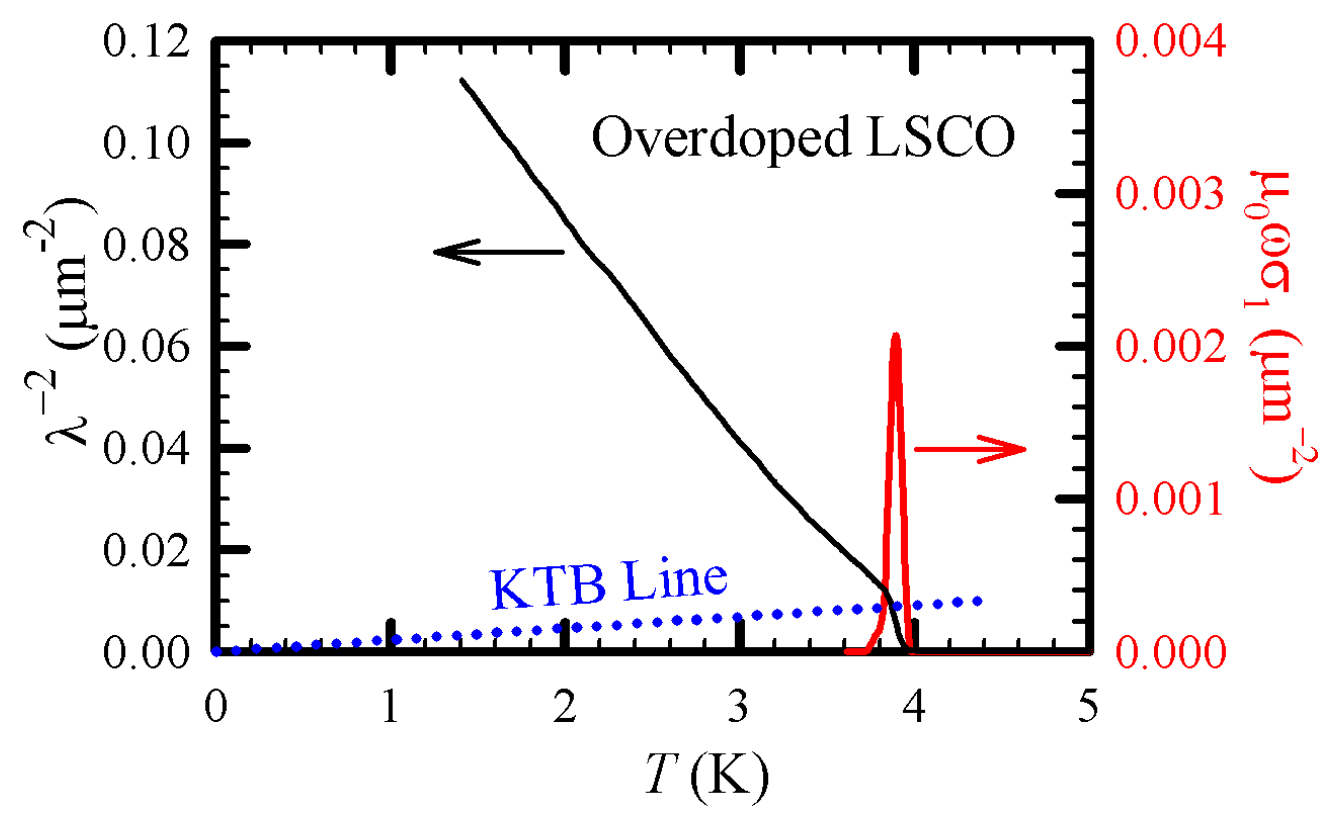

FIG. 3 (color online). $\lambda^{-2}(T)$ (black curve) and $\mu_0\omega\sigma_1(T)$ (red peak) for an overdoped LSCO film very close to the QPT. The KTB line (blue dotted) is calculated assuming the film fluctuates as a single 2D entity.

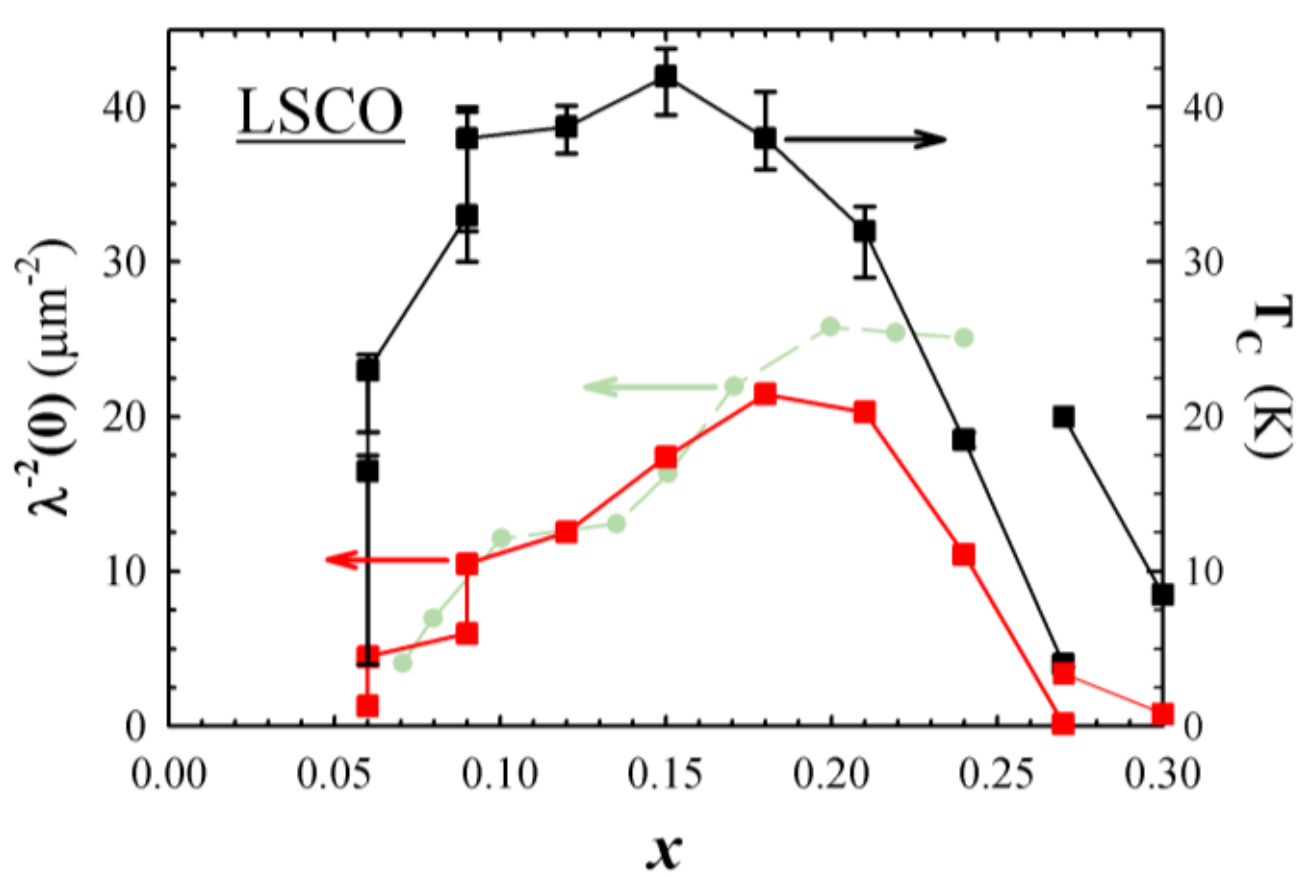

FIG. 4 (color online). $T_c$ (black squares) and $\lambda^{-2}(0)$ (red squares) vs. *x* for LSCO films; $\lambda^{-2}(0)$ vs. *x* for LSCO powders (green dots) [18].

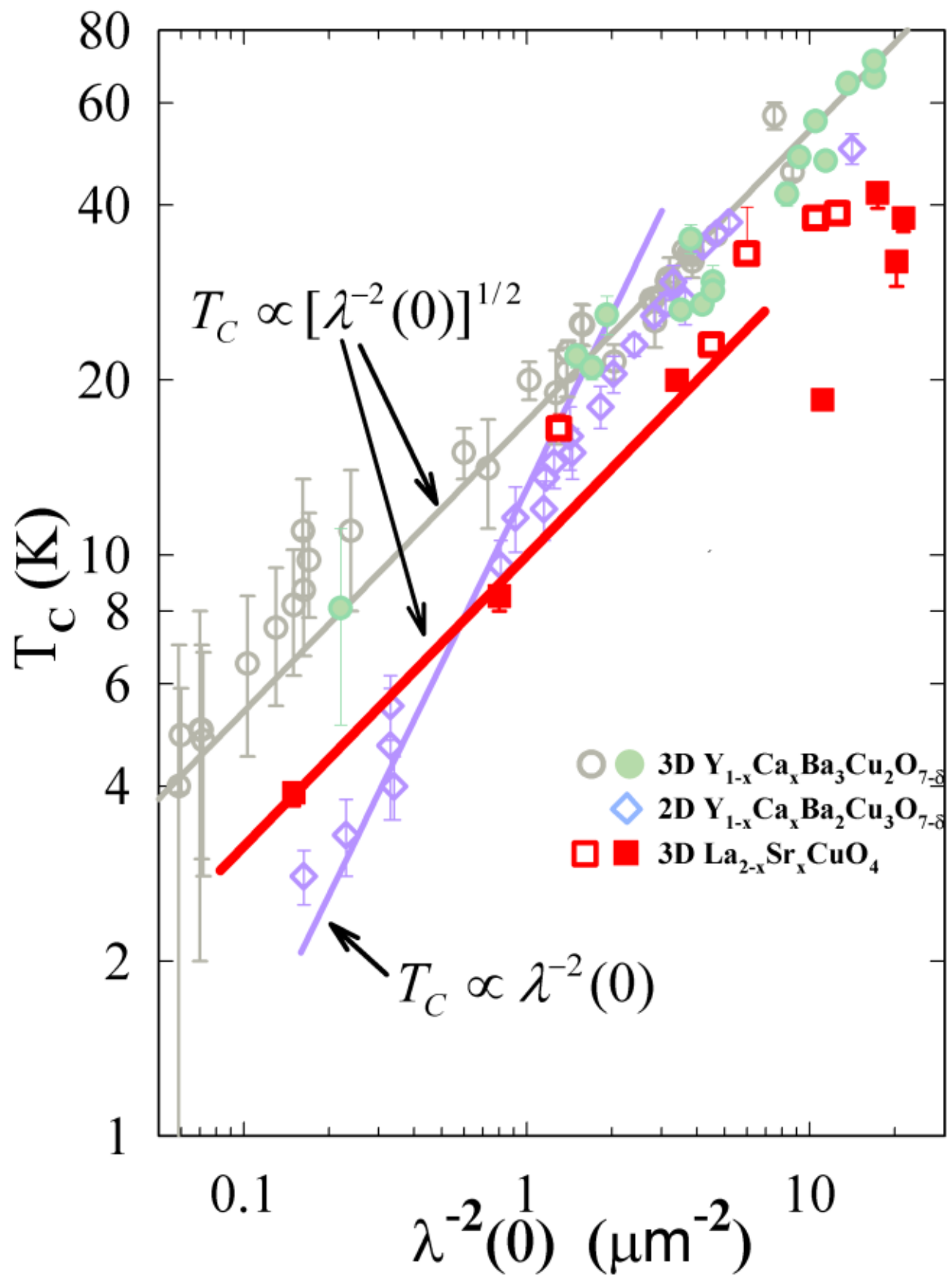

FIG. 5 (color online). $T_c$ vs. $\lambda^{-2}(0)$ for under- and overdoped LSCO films (open and filled red squares, respectively). Also shown are data for 40 unit-cell-thick YBCO [8] (filled green circles) and Ca-doped YBCO films [10] (open gray circles), and thin underdoped Ca-YBCO films [10] (open blue diamonds), which scale with 3D (2D) exponent α ≈ ½ (α = 1). Red line (α = ½) passes through the overdoped LSCO data.